# Visualizing intercalation effects in 2D materials using AFM based techniques

*Karmen Kapustić[+,1], Cosme G. Ayani[+,1], Borna Pielić[2], Kateřina Plevová[3], Šimun Mandić[1], Iva Šrut Rakić*[1]*

[1]Center for Advanced Laser Techniques, Institute of Physics, Bijenička Cesta 46, 10000, Zagreb, Croatia

[2]Department Physik, Universität Siegen, Walter-Flex-Str. 3, 57068 Siegen, Germany.

[3]Materials Science and Testing of Polymers, Department of Polymer Engineering and Science, Montanuniversitaet Leoben, 8700 Leoben, Austria

AUTHOR INFORMATION

[+]These authors contributed equally

**Corresponding Author**

*isrut@ifs.hr

Intercalation of two-dimensional materials, particularly transition metal dichalcogenides, is a non-invasive way to modify electronic, optical and structural properties of these materials. However, research of these atomic-scale phenomena usually relies on using Ultra-High Vacuum techniques which is time consuming, expensive and spatially limited. Here we utilize Atomic Force Microscopy (AFM) based techniques to visualize local structural and electronic changes of the $MoS_2$/graphene/Ir(111), caused by sulfur intercalation. AFM topography reveals structural changes, while phase imaging and mechanical measurements show reduced Young's modulus and adhesion. Kelvin Probe Force Microscopy highlights variations in surface potential and work function, aligning with intercalation signatures, while Photo-induced Force Microscopy detects enhanced optical response in intercalated regions. These results demonstrate the efficacy of AFM-based techniques in mapping intercalation, offering insights into tailoring 2D materials' electronic and optical properties. This work underscores the potential of AFM techniques for advanced material characterization and the development of 2D material applications.

**TOC GRAPHICS**

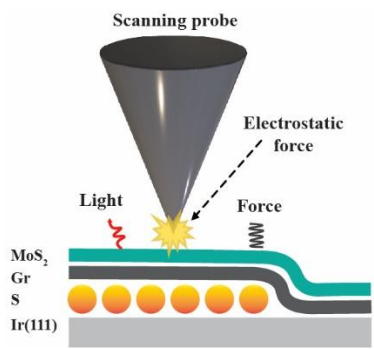

Two dimensional (2D) materials, especially transition metal dichalcogenides (TMDs), exhibit remarkable physical properties such as 2D superconductivity, correlated states, magnetism, and tunable bandgaps.[1–3] The latter, ability to change its band structure via functionalization, strain, substrate interaction, heterostructure assembly or intercalation is particularly important for a myriad of potential applications.[3–10] Intercalation of TMDs, especially, has emerged as a powerful approach to modify their properties as it can induce strain, phase transitions, change the doping, electronic structure and optical response, induce antiferromagnetic and ferromagnetic response and even cause an emergence of superconducting phase. [3,11–19] Additional appeal of intercalation is that it irreversible, nondestructive, and offers substantially greater range of chemical potential shift than gating[11,20]. From the 2D TMD family certainly the most studied one is molybdenum disulfide, $MoS_2$, due to its stability, flexibility and high charge mobility.[21–26] Intercalation and self-intercalation (intercalation of native atoms) also has interesting effects on monolayer $MoS_2$ as it was shown that it can induce metal-insulator transition, induce severe lattice changes and influence $MoS_2$-substrate interactions.[11–13]

TMDs are most commonly grown in CVD furnaces on atmospheric conditions, however growth in ultra-high vacuum (UHV), and in particular using molecular beam epitaxy (MBE), results in significantly better quality of 2D material displayed through low density of defects, pure crystal phases, and absence of adsorbents.[2,27,28] UHV growth also allows for an in situ measurements of crystal and electronic structure using wide range of techniques such as scanning tunneling microscopy (STM) and spectroscopy (STS) or angle resolved photoemission (ARPES). These techniques can offer an in-depth study; however, they are either good for local characterization at the nanoscale or global averaging at the microscale. This puts limitations on using them to study the inhomogeneous intercalation in TMDs, which can locally alter physical properties, and can easily be missed by many common UHV techniques. Furthermore, these techniques also require conductive and extremely flat samples, thus excluding insulators as well as samples grown in ambient conditions.

Here we show a visualization of local intercalation across various scales by exploiting structural, electrical and optical modes of atomic force microscopy (AFM). The adequate system to study the intercalation is $MoS_2$/Gr/Ir(111), because the $MoS_2$ shows good alignment on Gr/Ir, both Gr and $MoS_2$ are single layer crystals, and has naturally occurring self-intercalation which can easily be controlled.[11,12,29] On the other hand self-intercalation pathway and related effects on $MoS_2$ are up to now not well understood. Our sample is grown in UHV, characterized by STM and therefore used for comprehensive ex-situ AFM characterization. We show that local variation in $MoS_2$ growth creates two types of regions: (R1) merged $MoS_2$ islands surrounded by larger areas of uncovered graphene, and (R2) isolated, well defined, uniformly distributed $MoS_2$ islands. We correlate the observed changes with the vicinity of graphene wrinkles that enhance sulfur intercalation, resulting in weaker interaction between $MoS_2$ and Gr.[12] The region (R1), present around graphene wrinkles, show noticeable contrast in AFM phase image confirmed quantitatively with local drop in Young's modulus in comparison with region (R2). The differences in those two regions can also be distinguished by Kelvin probe force microscopy (KPFM), where we measure different work function, and more importantly, provide visual evidence that wrinkles present pathways for S intercalation. Finally, we employ a Photo-induced force microscopy (PiFM) to demonstrate its ability to map-out specific intercalated regions of our studied 2D heterostructure and show how intercalation increases measured optical signal. Our results show the versatility of AFM based techniques to visualize atomic-scale phenomena in 2D materials and highlights the importance to study MBE grown materials by both nanometer and micrometer scale imaging techniques.

We start with in-situ characterization of MoS$_2$/Gr/Ir(111) grown by two-step MBE process detailed in Experimental methods section. Low energy electron diffraction (LEED) patterns of graphene on Ir(111) before and after MoS$_2$ growth is shown in Figure 1a. Diffraction spots arising from Gr and Ir(111) are present on both images, while new spots, corresponding to MoS$_2$, appear on the second pattern. MoS$_2$ is well aligned to the Ir(111) first order spots with ± 5° distribution. Additional spots from intercalants or adatoms are not visible. Figures 1b and c show STM topography images of two different regions on the sample displaying apparent differences in morphologies. Namely, 1b shows mostly well defined, separate MoS$_2$ islands, while on 1c individual segments have aggregated and merged. We attribute the stripes, visible on both images, to the low coverage S intercalation below graphene,[30] however, area in 1c has ~9 times higher density of these stripes visible on graphene.

Figure 1d, recorded using AFM, shows a larger area next to wrinkles, comprising both distinct regions visually separated by dashed lines: Region R1 is close to the wrinkle network where MoS$_2$ islands are merged in dendritic structures, similar as in Figure 1c, and graphene appears adsorbate-free. The region R2 is distant from wrinkles, MoS$_2$ islands are compact, separated, homogeneously distributed and have higher coverage, similar as in Figure 1b (cf. also Supplementary Figure 1). Graphene on region R2 appears covered with adsorbates, likely from exposure to atmospheric conditions since UHV STM measurements did not show such contamination. Observed differences in MoS$_2$ morphology can be correlated to local S intercalation between graphene and Ir(111). Pielić et al.[12] have shown that self-intercalation of S weakens the interaction between MoS$_2$ and the substrate which results in high mobility of MoS$_2$ islands during growth. Individual islands then

coalesce into complex multi-domain structures. This is further elucidated by our STM data where we see increase in sulfur-related stripe density in regions of merged $MoS_2$ (Figure 1c). Resting on these previously published works of $MoS_2$ grown on gr/S/Ir(111)[12] and behavior of S intercalated graphene on Ir(111)[30] we can conclude that R1 regions next to wrinkles are in fact intercalated with S. We note that the concentration of intercalated S is below the detection threshold for averaging techniques such as LEED or Raman spectroscopy (cf. Supplementary Figure 2).

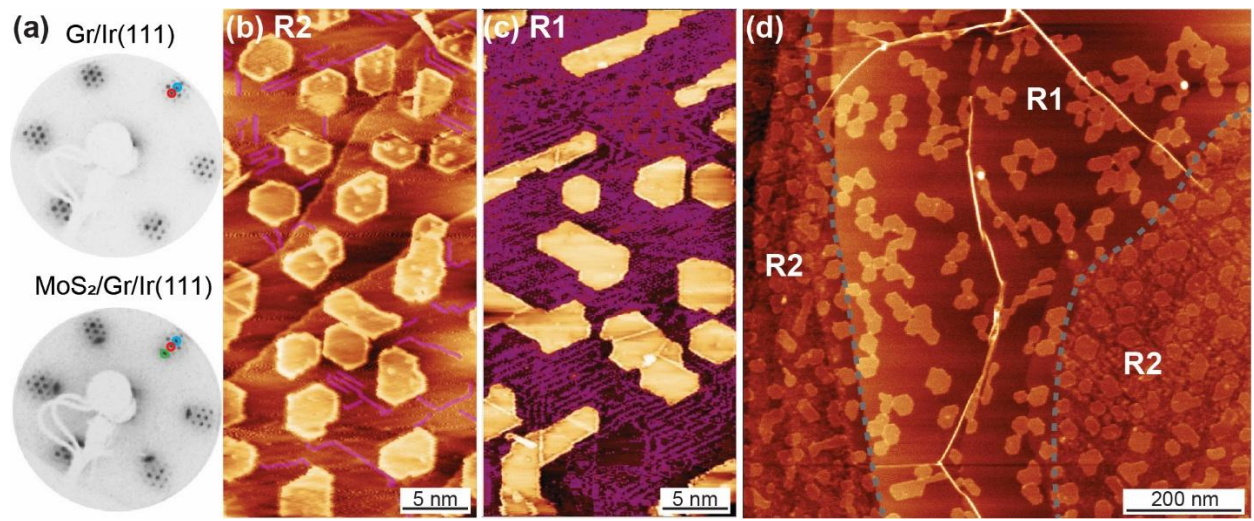

Figure 1. a) LEED image before and after $MoS_2$ growth on graphene on Ir(111). Intensity corresponding to graphene, Ir(111) and $MoS_2$ are marked with blue, red and green circles respectively. STM topography on two different regions on the $MoS_2$/Gr/Ir(111) sample where we display the extruded stripes in purple color: b) stripes cover ~6.2 % of bare graphene surface in image, and $MoS_2$ islands coverage is ~34.4%, c) stripes coverage is ~54.8 %, and $MoS_2$ island coverage is ~31.7%. d) Larger AFM image of the $MoS_2$/Gr/Ir(111) sample shows region around graphene wrinkles. LEED and STM data were recorded in-situ, while the AFM image is taken ex-situ, showing that the exposure to the air had no apparent effect on the morphology.

Large scale AFM (Figure 2a) topography (left) and phase (right) images reveal that, apart from the expected Ir steps and graphene wrinkles, we observe a strong change in phase contrast (blue arrows) corresponding to areas next to graphene wrinkles. Such local change in phase contrast i.e. phase shift is the first and most obvious indication that the sample is inhomogeneous. Phase gives information on the change in tip-sample interaction caused by a change in sample mechanical properties (adhesion, friction and stiffness).[31] This data can often be hidden in the simultaneously obtained topography channel. To gain a better understanding on the origin of the observed phase changes we employ a Quantitative imaging (QI) technique of our AFM. This is a force-curve based method which uses a special tip movement algorithm that measures a real and complete force-distance curve at every pixel of the image and gives all information about the local tip-sample interaction with high spatial resolution. This means that this method, unlike tapping or contact AFM modes, avoids lateral forces, allows for sub 50 pN forces to be applied and precisely maintained and most importantly gives quantitative measurements of e.g. adhesion force and Young's modulus along with the high-resolution topography images. Our QI measurements of the area containing two characteristic regions R1 and R2 are presented in Figure 2b showing measured topography, adhesion and Young's modulus. From these images we calculate that there is an average 1.1 nN decrease in adhesion and an average 50 MPa decrease in Young's modulus on the S intercalated areas. Decrease in Young's modulus is strictly observed on bare graphene of region R1, while the adhesion changes on a much larger scale and better outlines intercalated area (cf. Supplementary Figure 3), making it a good tool for identification of intercalation. Furthermore, adhesion channel also clearly shows positions of $MoS_2$ islands. Based on these measurements we conclude that the measured phase shift is combined result of local reduction in both adhesion and stiffness due to S intercalation, with former being the main contributing factor to the phase change.

Thus, both phase and adhesion modes can be used for quick and easy visualization of local intercalation.

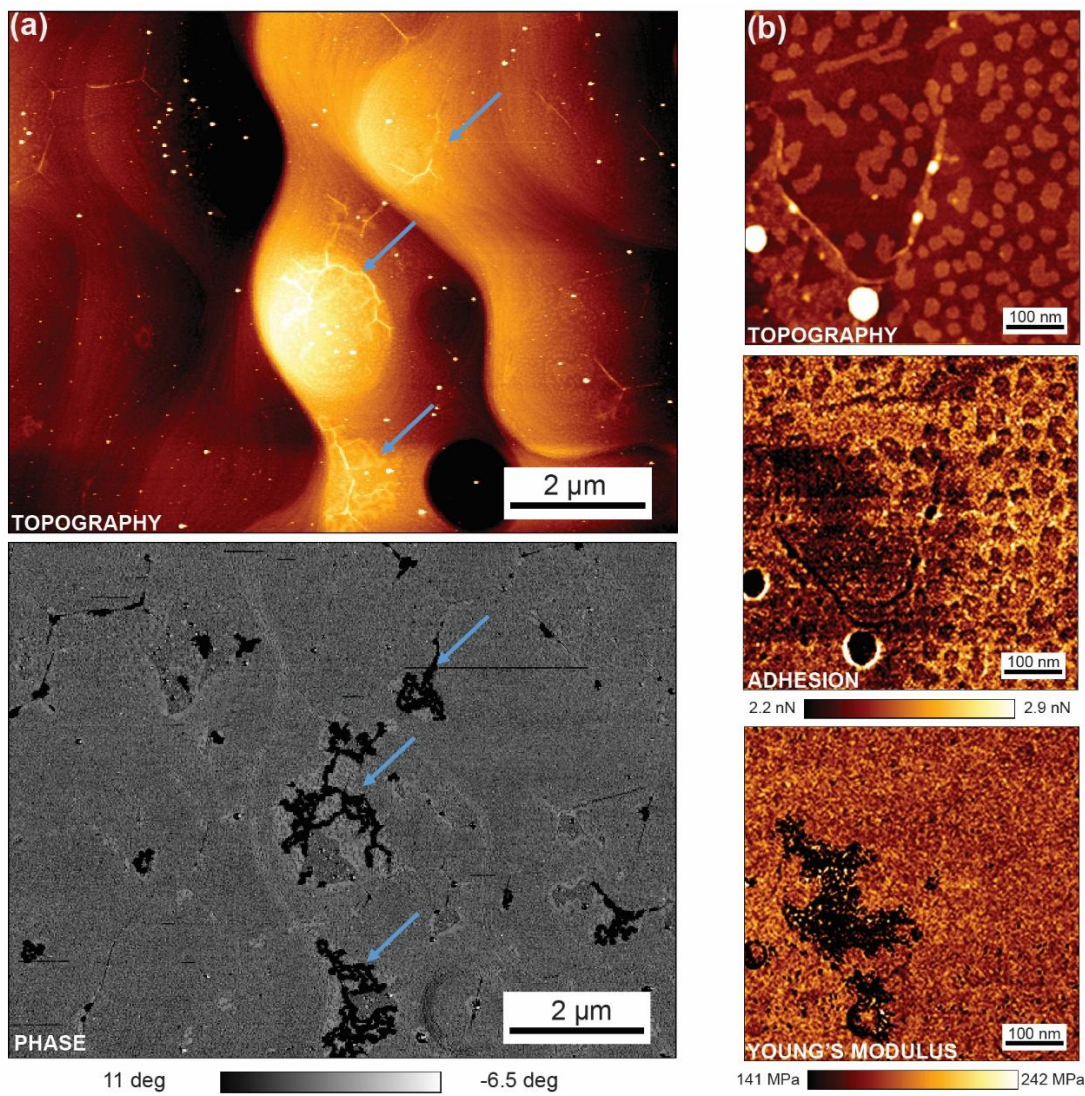

Figure 2. a) Large scale AFM image showing topography on the left and corresponding phase channel on the right. b) Topography (up), adhesion (middle) and Young's modulus (down) in a vicinity of a wrinkle, recorded by QI scanning mode. Young's modulus was extracted by fitting the measured force-distance curves (cf. Experimental Methods section). This approach is contingent to the assumed tip geometry and may lead to deviations from the true material values.

The work function is a surface property given by the specific crystal facet contributed by ~5 outermost atomic layers[32], and highly sensitive to changes in surface chemistry. Our previous work showed that intercalation can drastically change graphene doping [33,34] as well as $MoS_2$ band gap size and electronic band positions[12] so one would expect to see changes in work function as well. With that in mind we conducted an extensive KPFM measurements on our system, presented in Figure 3. On large scale images KPFM signal shows significant variation of surface potential. What is immediately obvious is the outline of the wrinkle network (cf. Figure 3a) which has enhanced surface potential. Wrinkles are accompanied by a 42 ± 1 mV increase of surface potential on a larger area in the wrinkle vicinity as can be seen in Figures 3b and c. Based on the $MoS_2$ islands morphology we conclude that those areas are the R1 regions intercalated with S. On both R1 and R2 regions $MoS_2$ islands show a surface potential decrease of 19 ± 2 mV compared to graphene and appear as dark features in KPFM. From Figure 3c it is evident that the surface potential for both graphene and $MoS_2$ changes when going from intercalated to non-intercalated regions. Calibrated KPFM measurements (Supplementary Figure 4) reveal that the graphene's work function drops from 4.935 ± 0.007 eV to 4.872 ± 0.008 eV going from pristine (R2) to intercalated (R1) area, while the $MoS_2$ WF drops from 4.954 ± 0.007 eV to 4.900 ± 0.008 eV. Please note that while surface potential upshifts for intercalated regions, the WF actually downshifts, due to specific tip WF and our setup geometry (see Supplementary information for more detail). We attribute the change in WFs largely to the S doping related changes in the electronic band structure of both graphene and $MoS_2$[12], even though a minor, localized contribution, likely comes from the surface potential variation due to wrinkles themselves (cf. Supplementary Figure 5). Measured WF of unintercalated graphene and $MoS_2$ in R2 regions is slightly higher than the values measured in UHV conditions which can be attributed to atmospheric contamination.[35–38] While atmospheric contamination can affect the WF it cannot fully account for the observed differences in the KPFM signal between R1 vs. R2, as adsorbates generally induce significantly weaker doping effects in 2D materials compared to intercalation.[34,39] Measured changes in KPFM signal caused by intercalation are more apparent than changes in surface morphology, especially on large scale images. Therefore, KPFM can be used as magnifying tool to detect intercalation and distinguish scale of intercalated regions.

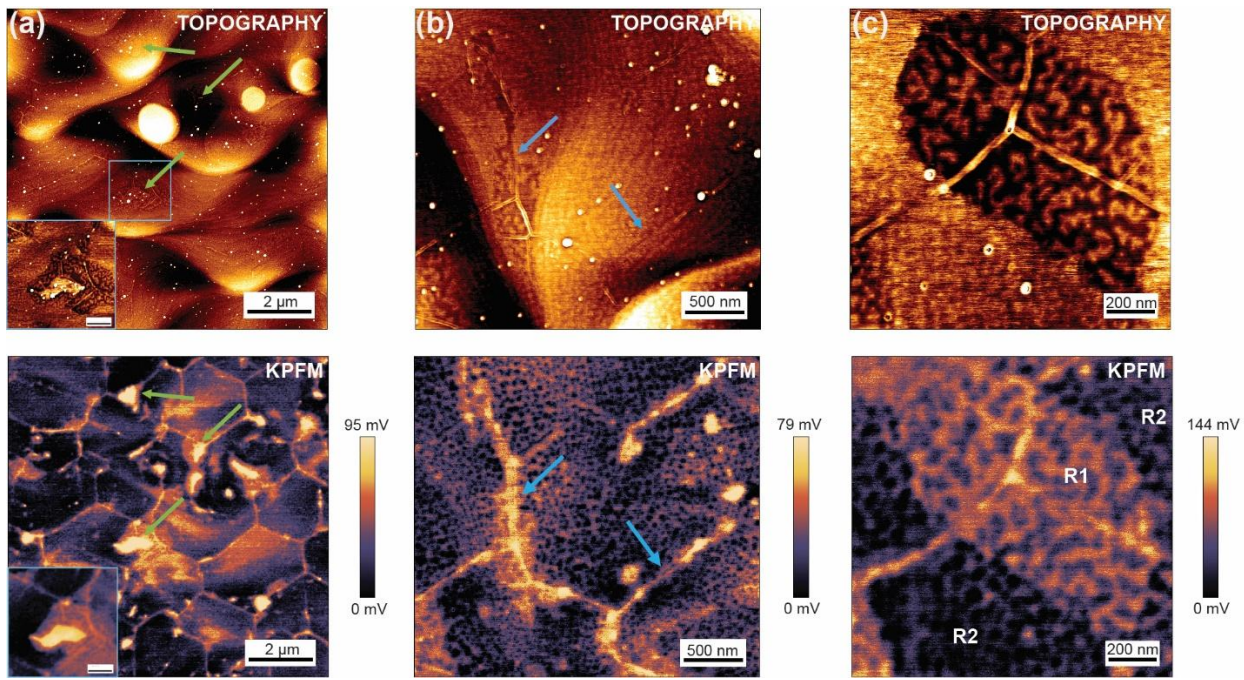

Figure 3. (a)-(c) AFM topographs (top panel) and corresponding KPFM images (bottom panel). Scale bar in all KPFM images was offset to 0 mV. Insets in figure (a) show a zoomed-in intercalated region, marked by blue square in topography panel, in a vicinity of a wrinkle. Scale bar in inset is 500 nm. The brightest areas in KPFM marked by green arrows are either residue of synthesis material or effect of oxidation due to air exposure. Blue arrows in (b) mark positions of missing wrinkles but with a same signature in KPFM as regular wrinkles. KPFM of the area imaged in (c) clearly shows how surface potential changes going from R1 to R2.

Interestingly, signs of wrinkle networks can be seen in KPFM data even when wrinkles itself are no longer present in topography (cf. Figure 3b blue arrows). As bare graphene on Ir(111) forms a complete network of wrinkles arranged in a honeycomb mesh,[40] the missing wrinkles are likely relaxed, though the mechanism of relaxation is currently unclear. However, visible signature of former wrinkle position in KPFM signal suggest presence of S intercalation localized around wrinkles. Petrović et al.[34] found that graphene wrinkles, and especially wrinkle crossing, have defects which act as penetration sites for atoms, while delaminated wrinkles serve as an atom conduit and a starting point for intercalation. While not all intercalant has the same intercalation mechanism, the localized presence of intercalated S next to wrinkles and in particular wrinkle crossings leads us to hypothesize that similar to Cs intercalation[34] wrinkles present intercalation pathways for sulfur. So far, intercalation pathway for S was only speculated to be wrinkle related,[30] but not experimentally confirmed, however more in-depth work needs to be done to fully understand atomic-level mechanism of S intercalation.

Optical properties of 2D materials often change hand-in-hand with electronic band structure. For that reason, we preform Photo induced force microscopy (PiFM) which measures an optical response of a sample in a tip-enhanced broadband infrared illumination. As it is an AFM based technique it gives topography image, together with optical response on a specific wave number, but can also be used to acquire local adsorption spectra consistent with Fourier transform infrared spectroscopy (FTIR)[41]. In Figure 4 we show data using 1270 cm$^{-1}$ for PiFM imaging as the accompanied spectra showed enhanced feature at that wavenumber. Large-scale topography image in Figure. 4a reveals wrinkles and accompanying intercalated R1 regions next to wrinkles with corresponding characteristic dendritic $MoS_2$ morphology. PiFM map of that same region shows intensity increase over the region R1. Zoom-in to an interface region between intercalated (R1) and non-intercalated (R2) areas shown in Figure 4 b confirms the optical response intensity increase by an average of 0.3 mV indicating adsorption increase. Spectra equivalent to adsorption FTIR spectroscopy (Figure 4c, Supplementary Figure 6), taken at different positions on the sample, reveal several distinct peaks, most notably at 1465 cm$^{-1}$ and 1263 cm$^{-1}$ as well as several smaller peaks. Analysis of the spectra reveals that, rather than distinct peak position shifts, we have the increase of the entire spectra intensity on intercalated graphene in R1 regions by up to 23% compared to non-intercalated graphene in R2 regions. Since we observe the same spectral features on both $MoS_2$ islands and graphene (cf. Supplementary Figure 6), and given how $MoS_2$ has a large optical gap and no active infra-red (IR) modes, we conclude that the spectral signal comes from graphene and becomes increased by S intercalation. Overlaying a non-IR active layer over an active one is expected to result in redshift of active modes, as well as decrease in peak intensity.[42] We notice some shifting of the prominent peaks going from $MoS_2$ to graphene, however this is of the order of the fitting error and thus we cannot claim any change in these specific vibrations. Since $MoS_2$ is not IR active, individual $MoS_2$ islands appear as dark features however they have bright edges as it can be seen on Figure 4d. We attribute the intensity increase on island edges to their metallic nature[43,44].

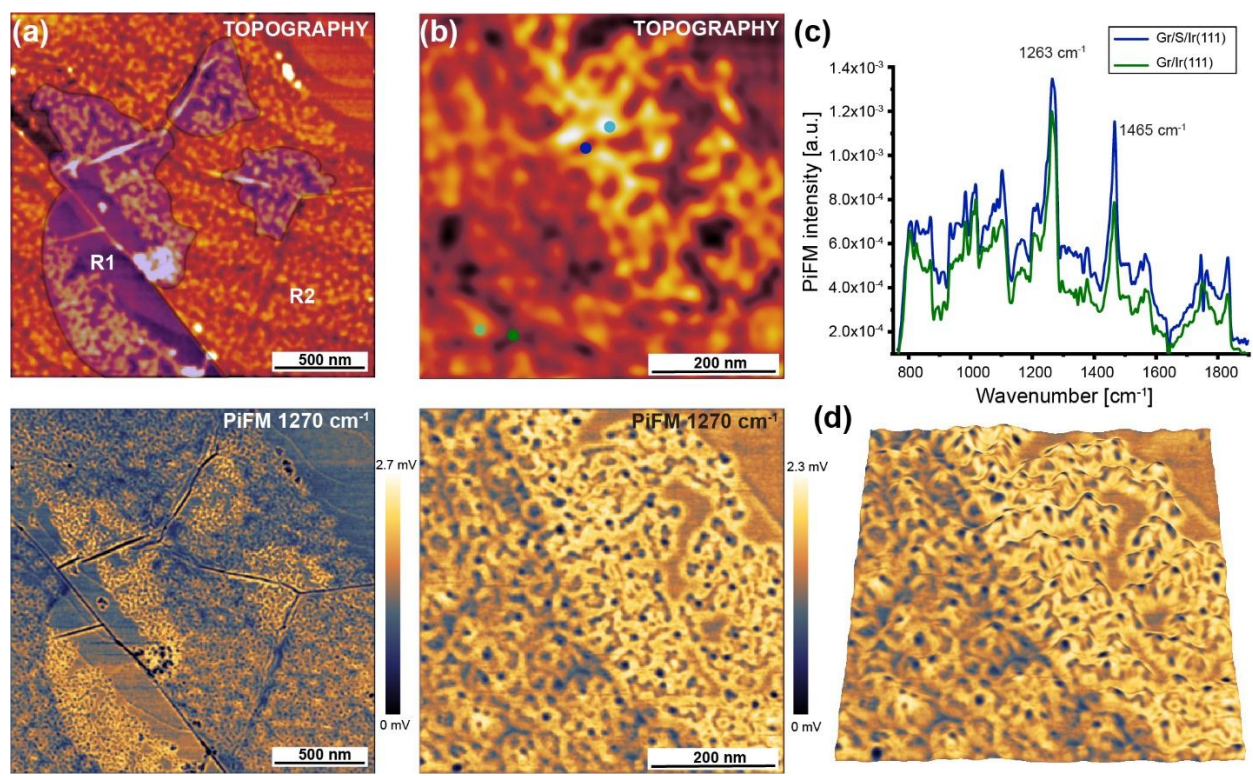

Figure 4. PiFM measurements. (a) Topography and PiFM signal of a large-scale image next to wrinkle. Areas shaded purple correspond to the regions R1 with increased optical signal in PiFM channel. (b) A topography and PiFM of a region showing intercalated R1 and non-intercalated R2 regions. (c) Full nano-FTIR spectra of intercalated (blue) and non-intercalated (green) graphene corresponding to positions marked in (b). Spectra taken over MoS$_2$ are presented in Supplementary Figure 3. (d) A 3D view of a topography in (b) overlaid with the PiFM signal.

While it is evident that intercalation leads to an overall increase of PiFM intensity allowing us the ability to use this technique as another intercalation visualization method, we can only speculate on the origin of this increase. Based on similarity in electronegativity of S and O,[45] S is expected to *p*-dope graphene. In that case larger charge density at the Fermi level presents higher probability of intraband electron-hole pair creation upon illumination, as well as an enhanced electron-phonon coupling due to the used energies of PiFM (~0.1 - 0.2 eV) falling within the phonon range energies of graphene (0-0.2 eV).[46] This, together with previously reported inverse relation between IR phonon and work function enhancement[47], that is also visible in our measurements, as well as broad range IR measurements on CVD grown graphene[48] indicates that the doping is likely reason for the IR phonon enhancement observed in our PiFM measurements. Still more comprehensive evaluation of the observed spectra and accompanying changes should be implemented in the future. These findings present compelling evidence that the IR phonon response can be used to map out intercalation.

This study demonstrates the effectiveness of AFM based techniques, including Kelvin Probe Force Microscopy, Quantitative imaging and Photo-induced Force Microscopy in visualizing and analyzing intercalation in two-dimensional materials, using a case study of S intercalated $MoS_2$ on Gr/Ir(111) substrate. Our findings show that S intercalation significantly impacts the morphology, mechanical properties, electronic structure and optical response of the $MoS_2$/graphene system. The local variations in phase contrast, Young's modulus, adhesion, and work function provide a detailed understanding of intercalation effects. Notably, KPFM measurements allowed us to reveal that wrinkles are an intercalation pathway for S intercalation. Additionally, the increased optical response intensity observed through PiFM further maps-out the presence of intercalation. These results highlight the versatility of AFM-based techniques for studying local variation in the properties of 2D materials, paving the way for future advancements in material science and nanotechnology applications.

Experimental Methods

**Sample preparation.** Ir(111) single crystal was used in all measurements (99.99% purity and orientation accuracy better than 0.1°). The substrate was cleaned by standard UHV techniques of Ar$^+$ ion sputtering and annealing.[49] Gr on Ir(111) is prepared in UHV conditions by applying the cycle of temperature-programmed growth (TPG), 6 L of ethylene dosed at room temperature (RT) and post-annealed to 1270 K, and the chemical vapor deposition (CVD), 140 L of ethylene dosed at 1130 K.[28] Graphene quality was checked by LEED (cf. Figure 1a) and aligned R0° Gr/Ir(111) is confirmed. MoS$_2$ islands are grown by MBE two-step growth method[27]. In the first step Mo is deposited to Gr/Ir(111) at RT with continuous exposure to sulfur pressure (6×10$^{-9}$ mbar) provided by pyrite heating. In the second step the sample was annealed to 1000 K at the same sulfur pressure. This procedure is optimized to minimize naturally occurring self-intercalation of sulfur atoms, however, sulfur can still be present even in low intercalation regime where averaging techniques such as LEED produce no observable signal.

**STM.** The STM characterization was performed on a SPECS Aarhus STM using etched tungsten tip. The data were collected at RT, with the STM tip grounded and the sample put to a bias voltage. STM data analysis was performed in the WSxM software.[50]

**AFM.** Atomic force microscopy (AFM) topography images were acquired in atmospheric conditions using JPK NanoWizard 4 AFM Ultra speed in AC mode. AFM probes were obtained from Bruker (FESP – V2 with a radius of curvature of < 10 nm, nominal spring constant of 2.8 N/m and a nominal resonant frequency of 75 kHz). Young's modulus, adhesion and corresponding topography images were acquired in QI (Quantitative imaging) mode using the same probes as for AC mode. Adhesion is measured directly, while in order to get Young's modulus we applied the fitting process using Derjaguin–Muller–Toporov (DMT) model with a triangular pyramid tip shape. Surface potential/work function images were acquired in KPFM mode using Pt/Ir coated probes from NanoWorld (NCHPT with a radius of curvature of < 25 nm, nominal spring constant of 42 N/m and a nominal resonant frequency of 320 kHz) with a grounded sample and modulation amplitude of 1.2 V. Probes were calibrated on a freshly cleaved HOPG with a known WF of 4.48 eV[36] to determine the probe and subsequently sample WF. All images were processed using JPK Data processing, WSxM[50] and Gwyddion software.[51] All KPFM data presented in the main text have a Z scale set to 0 V. Average surface potential was determined by analyzing histograms of specific region and fitting a Gauss curve, while for the error we use RMS roughness of the analyzed surface as it better reflects variations in surface potential than the Gauss fit error.

**PiFM.** The Photo-induced force microscopy (PiFM) was performed using a VistaScope 75 (Molecular Vista, USA) in tapping mode with an amplitude of 2 nm. As a probe, NCH-PtIR tip (nominal force constant 42 N/m, nominal resonance frequency 250 kHz) and NCHR-Au tip (nominal force constant 42 N/m, nominal resonance frequency 270 kHz) by Molecular Vista were used. The source of IR light was a QCL laser with a working range of 770-1850 $cm^{-1}$.

The PiFM scans were recorded using 3-5% of the laser power at different wavenumbers (856, 871, 959, 998, 1008, 1096, 1105, 1153, 1161, 1207, 1279, 1324, 1344, and 1680 $cm^{-1}$) during scanning. PiFM, topography, and phase images were recorded simultaneously. The full spectra in the range of 770-1850 $cm^{-1}$ were measured in side-band mode with 4% of the laser power within 30 seconds.

**Raman spectroscopy**. Raman spectra were taken with a commercial Renishaw inVia confocal microscope at ambient conditions and room temperature. It is equipped with a 532 nm (2.33 eV) continuous wave laser source. Measurements were taken with a 50× objective (NA = 0.50) and a grating with a 2400 $mm^{-1}$ constant.

ASSOCIATED CONTENT

**Supporting Information**.

Supplementary Information.pdf includes additional Raman spectroscopy, STM, AFM and nanoFTIR measurements.

AUTHOR INFORMATION

**Notes**

[+] these authors contributed equally

The authors declare no competing financial interests.

ACKNOWLEDGMENT

This research was supported by the project *Sustav za rast i naprezanje materijala u vakuumskim uvjetima (STRAIN - mat)* financed by the European Union – NextGenerationEU through the National Recovery and Resilience Plan 2021-2026 (NRPP). B.P. acknowledges financial support by Marie Skłodowska-Curie Actions (MSCA) Postdoctoral Fellowship (project 101107288, 2D-InTune, B.P.). C.G.A. acknowledges financial support by the Croatian Science Foundation, Grants No. UIP-2020-02-8891 and MOBDOL-2023-08-2165. I.Š.R. acknowledges financial support by bilateral Croatian-Austrian project funded by Croatian Ministry of Science and Education and the Centre for International Cooperation and Mobility (ICM) of the Austrian Agency for International Cooperation in Education and Research (OeAD-GmbH) under project HR 02/2020.